\documentclass[conference,a4paper]{IEEEtran}
\ifCLASSINFOpdf
\usepackage{graphicx}
\usepackage{tikz}
\usetikzlibrary{shapes.geometric, arrows, positioning, decorations.pathreplacing}
\usepackage{float}
\usepackage{multirow}
\begin{document}

\title{Leveraging LLMs for Mental Health: Detection and Recommendations from Social Discussions}

\author{
\IEEEauthorblockN{Vaishali Aggarwal}
\IEEEauthorblockA{\textit{TCS Research} \\
Delhi, India  \\
aggarwal.vaishali@tcs.com
}
\and
\IEEEauthorblockN{Sachin Thukral}
\IEEEauthorblockA{\textit{TCS Research} \\
Delhi, India \\
sachi.2@tcs.com}
\and
\IEEEauthorblockN{Krushil Patel}
\IEEEauthorblockA{\textit{IIT Kharagpur} \\
West Bengal, India \\
krushilpatel@kgpian.iitkgp.ac.in
}
\and
\IEEEauthorblockN{Arnab Chatterjee}
\IEEEauthorblockA{\textit{TCS Research} \\
Delhi, India \\
arnab.chatterjee4@tcs.com
}
}

\maketitle

\begin{abstract}

Textual data from social platforms captures various aspects of mental health through discussions around and across issues, while users reach out for help and others sympathize and offer support. We propose a comprehensive framework that leverages Natural Language Processing (NLP) and Generative AI techniques to identify and assess mental health disorders, detect their severity, and create recommendations for behavior change and therapeutic interventions based on users' posts on Reddit.

To classify the disorders, we use rule-based labeling methods as well as advanced pre-trained NLP models to extract nuanced semantic features from the data.
We fine-tune domain-adapted and generic pre-trained NLP models based on predictions from specialized Large Language Models (LLMs) to improve classification accuracy. Our hybrid approach combines the generalization capabilities of pre-trained models with the domain-specific insights captured by LLMs, providing an improved understanding of mental health discourse. Our findings highlight the strengths and limitations of each model, offering valuable insights into their practical applicability.

This research potentially facilitates early detection and personalized care to aid practitioners and aims to facilitate timely interventions and improve overall well-being, thereby contributing to the broader field of mental health surveillance and digital health analytics.
\end{abstract}


\section{Introduction}

Mental health is crucial for overall well-being, with early detection and intervention being vital. Mental illness affects millions worldwide and is a significant health burden, worsened by the COVID-19 pandemic~\cite{santomauro2021global}. Social media, especially Reddit, offers valuable insights into mental health through extensive user discussions around experiences and struggles. However, analyzing this data is challenging~\cite{de2014mental} due to informal language, slang, and emojis, which complicate traditional NLP techniques. 

Generative AI, particularly Large Language Models (LLMs), has advanced the analysis of unstructured mental health data but poses risks like misinformation and misdiagnosis. Ethical concerns, biases, and the need for human oversight are crucial. Ongoing efforts aim to improve LLMs while addressing these challenges.

Advanced pre-trained NLP models are essential for processing text -- they excel in classification, sentiment analysis, and semantic understanding. Learning from labeled data, they improve classification performance and effectively interpret nuanced language in social media discussions.

We propose a framework to leverage the knowledge of LLMs to:
(i) \textbf{Identify Mental Health Issues}: Utilize social media data to detect whether it is related to mental health, 
(ii) \textbf{Classify Disorders and Severity}: Accurately classify the disorders mentioned in the posts and assess their severity, and 
(iii) \textbf{Recommend Interventions}: Provide recommendations for  behavior changes and therapies based on the identified disorders and their severity.
 
This paper presents a framework that utilizes large-scale Reddit data on mental health, combining LLMs with advanced NLP models for a multi-label classification. This integration aims to enhance the precision and reliability of mental health analysis 
by addressing the complexities of social media text and bridging the gap between vast social media data and actionable mental health insights.

The research has the potential to advance mental health informatics by enabling more accurate identification of mental health disorders from social media, potentially facilitating earlier interventions and improving lives.

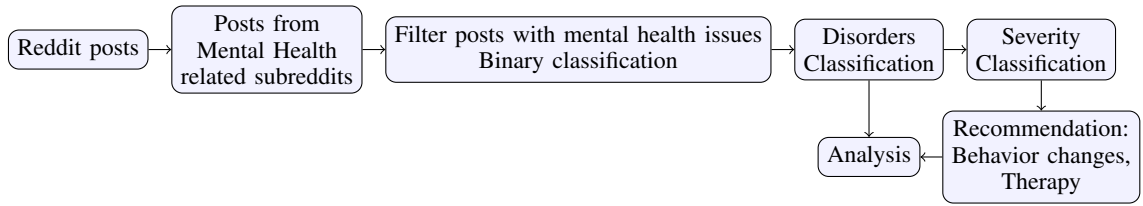
\begin{figure*}[!th]

   \resizebox{\textwidth}{!}{
   \centering 
    \begin{tikzpicture}[node distance=0.25cm and 0.15cm, every node/.style={fill=blue!05, draw, rounded corners, minimum height=0.5cm, minimum width=1.5cm, font = \small, align=center}]
        \tikzset{font=\small}

    
        \node (empty) [fill=none,draw=none]{};
        \node (reddit) [rectangle, minimum width = 0.25cm, right= of empty] {Reddit posts};
        \node (posts) [right=of reddit,xshift=0.15cm, minimum width=0.25cm] {Posts from\\ Mental Health \\related subreddits};

        \node (binary) [rectangle, right=of posts, xshift=0.15cm, minimum width = 0.25cm] {Filter posts with mental health issues\\ Binary classification};

        \node (disorders) [rectangle, right=of binary,xshift=0.15cm, minimum width =0.25cm] {Disorders\\ Classification};

        \node (severity) [rectangle, right=of disorders, xshift=0.15cm, minimum width =0.25cm] {Severity\\ Classification};

        \node (analysis) [rectangle, below=of disorders, yshift=-0.50cm, minimum width = 0.25cm] {Analysis};

        \node (recommendation) [rectangle, below=of severity, yshift=-0.15cm, minimum width=0.25cm] {Recommendation: \\ Behavior changes,\\ Therapy};

        \node (empty2) [fill=none,draw=none, right =of severity]{};
       

        \draw [->] (reddit) -- (posts);

        \draw [->] (posts) -- (binary);

        \draw [->] (binary) -- (disorders);

        \draw [->] (disorders) -- (severity);

        \draw [->] (disorders) -- (analysis);

        \draw [->] (severity) -- (recommendation);

        \draw [->] (recommendation) -- (analysis);

    \end{tikzpicture}
   }
    \caption{Framework for analyzing Reddit posts on mental health, with modules for disorders classification, severity detection, behavior change recommendation and therapy recommendation.}
    \label{fig:Pipeline}
\end{figure*}

\section{Related work}
Analyzing social media data for mental health diagnosis has become increasingly significant due to the wealth of information available on social media platforms like Twitter, Reddit, etc. Various methodologies developed to enhance the accuracy of identifying mental health conditions from user-generated content. Linguistic markers of depression examined from Twitter data~\cite{wolohan2018detecting} highlighted the potential of social media for mental health insights. Recent work on identifying stress factors that lead to mental health decline has demonstrated the effectiveness of NLP techniques in this domain~\cite{thukral2020identifying}. 
Various machine learning methods, particularly deep learning, have improved detection accuracy for mental health issues on social media~\cite{tadesse2019detection}. The "Dreaddit" dataset showcased NLP techniques for identifying stress-related content on Reddit through binary classification~\cite{turcan2019dreaddit}. Additionally, a gold standard dataset~\cite{sampath2022data} was developed for detecting depression levels using traditional algorithms and data augmentation in a multi-class framework. RoBERTa~\cite{RoBERTa} enhances BERT's performance through optimized pre-training, but our focus on mental health aligns better with domain-specific models like MentalRoBERTa~\cite{ji2021mentalbert}. ``Illuminate''~\cite{agrawal2024illuminate} leverage GPT-4 and Llama2 for depression detection and therapy recommendations through advanced prompt engineering. A study analyzing LLMs as therapists~\cite{chiu2024computational} suggests they exhibit low-quality therapy behaviors but also address client needs, indicating the need for further research to improve care quality.
 
\section{Dataset}
Reddit is a popular platform with active communities organized into subreddits. We extracted text content from posts in the subreddits \texttt{r/mentalhealth} and \texttt{r/depression} for the period of December 1, 2019 to November 30, 2020, totaling 334,197 posts using the Pushshift~\cite{Reddit_dump} API. Each post is treated as an independent entity, regardless of the user, to focus on content analysis. To validate our methodology, we selected a representative subset of 5,000 random posts from the dataset, ensuring it captures the essential characteristics for evaluating our approach.

\section{Methodology}
\label{sec:methodology}
Our proposed framework, outlined in Fig.~\ref{fig:Pipeline} for identifying mental health conditions from Reddit posts, involved several key steps, which are elaborated below.

\subsection*{Leveraging Large Language Models (LLMs)}
We used Llama3~\cite{dubey2024llama3herdmodels}, MentaLLaMA~\cite{Yang_2024}, and Samantha-Mistral for our analysis, exploring their potential for mental health condition identification and analysis. MentaLLaMA, based on LLaMA2 models, offers significant improvements in providing both accurate classifications and high-quality explanations for mental health analysis on social media. MentaLLaMA addresses the challenge of low interpretability in traditional mental health analysis methods by building the Interpretable Mental Health Instruction (IMHI) dataset.
Samantha-Mistral LLM, also known as Samantha-1.2-Mistral-7B\footnote{https://huggingface.co/cognitivecomputations/samantha-1.2-mistral-7b}, is an advanced language model developed to function as a caring and empathetic AI companion, and has been trained extensively in philosophy, psychology, and personal relationships, making it uniquely suited to engage in human-like interactions and offer support in these areas.

\subsection{Preprocessing using Binary Classification}
\label{subsec:preprocessing}
We employed various LLMs to filter out irrelevant content, thereby refining the dataset to include only posts that were genuinely related to mental health conditions. Each post was evaluated to determine whether it pertained to mental health, with the LLMs instructed to provide responses strictly in \textit{yes} or 
\textit{no} format. By leveraging the advanced capabilities of LLMs, we were able to efficiently and accurately classify the vast number of posts, ensuring the integrity and focus of our analysis on meaningful mental health-related content (see Table.~\ref{tab:binary}).

\begin{table}[h]
\caption{Preprocessing for mental health related posts}
	\label{tab:binary}
	\begin{center}
	    \begin{tabular}{|p{3.3cm}|r|r|r|}
     \hline
     \textbf{Model Name} & \textbf{Yes} & \textbf{No} & \textbf{Other} \\ 
     \hline
     Llama3 & 67.7\% & 27\% & 5.3\% \\ 
     \hline
     MentaLLaMA & 85.9\% & 12\% & 2.1\% \\ 
     \hline
     Samantha-Mistral & 58.6\% & 41.3\% & 0\% \\ 
     \hline
        \end{tabular}
	\end{center}
\end{table}

\subsection{Disorders Classification}
\label{subsecm:symptom}
\subsubsection{Data Labeling} For this study, we selected nine labels based on the Self-reported Mental Health Diagnoses (SMHD) dataset~\cite{cohan2018smhd} which contains dictionaries for nine distinct mental health disorders -- \textit{Attention-Deficit Hyperactivity Disorder (ADHD), Autism, Anxiety, Bipolar Disorder, Depression, Eating Disorder, Obsessive-Compulsive Disorder (OCD), Post-Traumatic Stress Disorder (PTSD)} and \textit{Schizophrenia}. This selection allows us to comprehensively cover a range of mental health issues within our analysis.

\begin{figure}[t]

   \resizebox{\linewidth}{!}{
   \centering 
    \begin{tikzpicture}[node distance=0.25cm and 0.15cm, every node/.style={fill=blue!05, draw, rounded corners, font = \tiny, align=center}]
        \tikzset{font=\tiny}

        \node (reddit) [rectangle, fill=none, draw=none, minimum height=0cm, minimum height=0cm] {\includegraphics[width=0.5cm]{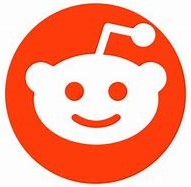}};
        \node (empty) [fill=none,draw=none]{};
        \node (dataset) [right=of reddit, fill=none, draw=none, minimum height=0cm, minimum width=0cm] {\includegraphics[width=0.5cm]{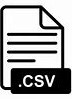}};
        \node (labelled) [rectangle, right=of dataset, xshift=1.7cm] {Labelled Data};
        \node (dictionary) [fill=none, draw=none, above=of labelled, yshift=0cm, xshift=-0.65cm, rotate=325] {Dictionary\\ based};
        \node (prompt) [fill=none, draw=none, below=of dataset, xshift=0.7cm,rotate=330,yshift=0.25cm] {Prompt};
        \node (response) [fill=none, draw=none, below=of labelled, yshift=0cm, xshift=-0.7cm, rotate=35] {Response};
        \node (SMHD) [rectangle, above=of dataset, xshift=1.5cm, minimum width = 0.25cm, fill=red!10] {SMHD \\dataset};
        \node (LLMs) [rectangle, below=of dataset, xshift=1.5cm, minimum width = 0.25cm, fill=red!10] {LLMs};
        \node (NLP) [rectangle, right=of labelled, minimum width=0.25cm] {Training of\\pre-trained\\NLP models};
        \node (results) [rectangle, right=of NLP, minimum width=0.5cm, xshift=0.1cm] {Classification results};
        \node (metrics) [rectangle, below=of results] {Performance Metrics};
        \node (empty2) [fill=none,draw=none, right =of LLMs]{};


        \draw [->] (reddit) -- (dataset);
        \draw [-] (dataset) -- (SMHD);
        \draw [-] (dataset) -- (LLMs);
        \draw[->] (SMHD) -- (labelled);
        \draw[->] (LLMs) -- (labelled);
        \draw [->] (labelled) -- (NLP);
        \draw[->] (NLP) -- (results);
        \draw[->] (results) -- (metrics);
    \end{tikzpicture}
}
    \caption{Framework for disorders classification for Reddit posts on mental health.}
    \label{fig:Pbline_symp}   
\end{figure}
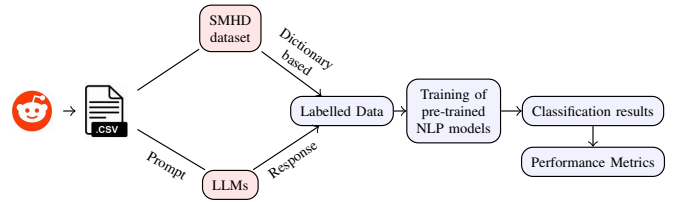

\subsubsection{Data Annotation}
We employed four different annotators to label Reddit posts for mental health disorders. These annotators included (i) dictionary-based labeling, (ii) Llama3, (iii) Samantha-Mistral, and (iv) MentaLLaMA. 
 
We used SMHD dataset dictionaries to label Reddit posts by matching terms linked to mental health disorders. Posts with terms from multiple dictionaries received multiple labels, while unmatched posts were labeled as \textit{None}. For the LLM based annotations, a suitable prompt classified posts with pre-defined labels for different disorders. The overall framework of disorders classification is shown in Fig.~\ref{fig:Pbline_symp}.

To ensure consistency, we filtered predictions to include only labels for the nine predefined mental health disorders. This multi-annotator approach allowed us to effectively compare traditional dictionary-based methods with advanced LLMs in identifying mental health disorders.
 
\subsubsection{Training Pre-trained Models}


We evaluated various pre-trained NLP models, including domain-specific ones like MentalRoBERTa~\cite{ji2021mentalbert}, and general models like DistilBERT~\cite{sanh2019distilbert}. Each model was fine-tuned and trained to classify posts based on labeled data from different annotators to improve accuracy. Input sequences were tokenized using respective pre-trained tokenizers, truncated, and padded to a maximum of 512 tokens. Both models were fine-tuned using the Binary Cross-Entropy loss function with the AdamW optimizer, optimizing them for identifying mental health disorders in Reddit posts.
 
\subsubsection{Model Evaluation} 
We evaluated the performance of both pre-trained NLP models and LLMs using standard metrics such as accuracy, precision, recall, and F1-score. For validation, a 5-fold cross-validation strategy was employed, training each fold for 10 epochs. We assessed the generalizability of our trained models on a new, unseen sample of Reddit posts related to mental health, comparing their predictions with dictionary-assigned labels and LLM predictions. This comprehensive evaluation framework aimed to improve the accuracy and robustness of mental health disorders identification, facilitating timely interventions and better well-being outcomes.  

\subsection{Severity Classification}
\label{subsecm:severity}
We utilized various LLMs to assess the severity level of each user's post, classifying them into three distinct categories: \textit{mild}, \textit{moderate}, and \textit{severe}. 

\subsection{Recommendation Module}
\label{subsecm:recom}
\subsubsection{Suitable Therapy}
This module leverages LLMs to provide personalized therapy suggestions based on the severity of each user post. LLMs identify suitable therapies and offer a confidence percentage for each, allowing for comparison and ensuring the recommendations are both appropriate and backed by a quantifiable measure.

\subsubsection{Behaviour Change}
The recommendation module also includes suggestions for behavior changes to improve mental well-being. For each user post, we generated tailored recommendations for behavior changes that could positively impact the user's mental health, crafted based on the specific context and content of the posts, ensuring that they are relevant and actionable. 

The above provides personalized behavior advice practical strategies to help users improve mental well-being, complementing therapy for a holistic mental health approach.

\section{Results}

In the rest of the paper, we demonstrate our results on a random sample of $5000$ posts.

\subsection{Preprocessing using Binary Classification}
\label{subsecr:preprocess}

The results corresponding to binary classification task for various LLMs are given in Table.~\ref{tab:binary}. 
Here, \textit{Other} denotes the cases where the LLM failed to perform the classification. We observed that 
MentaLLaMA, being a domain-specific model, classifies the majority of the data as being related to mental health disorders. 
The set of posts that are classified as ``YES'' in the above constitutes our final dataset for further analysis.

\subsection{Disorders Classification}
\label{subsecr:symptom}
During fine-tuning, we experimented with different learning rates for both DistilBERT and MentalRoBERTa, achieving the best accuracy with $3e-5$.

\begin{table*}[t]
\centering
\caption{NLP Model Evaluation}
\label{tab:NLP Models}
\begin{tabular}{|l|lllllllllc|}
\hline
\multicolumn{1}{|c|}{} & \multicolumn{10}{c|}{Models}                                                                                                                                                                                                   \\ \hline
\multicolumn{1}{|c|}{Annotators} & \multicolumn{5}{c|}{DistilBERT}                                                                                                 & \multicolumn{5}{c|}{Mental RoBERTa}                                                                            \\ \hline
                       & \multicolumn{1}{l|}{Tr. Acc. \%} & \multicolumn{1}{l|}{Val. Acc. \%} & \multicolumn{1}{l|}{Precision} & \multicolumn{1}{l|}{Recall} & \multicolumn{1}{l|}{F1 Score} & \multicolumn{1}{l|}{Tr. Acc. \%} & \multicolumn{1}{l|}{Val. Acc. \%} & \multicolumn{1}{l|}{Precision} & \multicolumn{1}{l|}{Recall} & F1 Score \\ \hline
                       Dictionary Matching & \multicolumn{1}{c|}{87.7} & \multicolumn{1}{c|}{86.25} & \multicolumn{1}{c|}{0.88} & \multicolumn{1}{c|}{0.87} & \multicolumn{1}{c|}{0.87} & \multicolumn{1}{c|}{84.16} & \multicolumn{1}{c|}{82.73} & \multicolumn{1}{c|}{0.83} & \multicolumn{1}{c|}{0.87} & 0.85 \\ \hline
                       Llama3 & \multicolumn{1}{c|}{95.8} & \multicolumn{1}{c|}{70.31} & \multicolumn{1}{c|}{0.88} & \multicolumn{1}{c|}{0.88} & \multicolumn{1}{c|}{0.88} & \multicolumn{1}{c|}{91.02} & \multicolumn{1}{c|}{69.42} & \multicolumn{1}{c|}{0.87} & \multicolumn{1}{c|}{0.91} &  0.88\\ \hline
                       MentaLLaMA & \multicolumn{1}{c|}{94.15} & \multicolumn{1}{c|}{59.21} & \multicolumn{1}{c|}{0.81} & \multicolumn{1}{c|}{0.85} & \multicolumn{1}{c|}{0.82} & \multicolumn{1}{c|}{92.34} & \multicolumn{1}{c|}{60.64} & \multicolumn{1}{c|}{0.78} & \multicolumn{1}{c|}{0.8} & 0.79 \\ \hline
                        Samantha-Mistral & \multicolumn{1}{c|}{83.47} & \multicolumn{1}{c|}{32.40} & \multicolumn{1}{c|}{0.71} & \multicolumn{1}{c|}{0.82} & \multicolumn{1}{c|}{0.75} & \multicolumn{1}{c|}{79.26} & \multicolumn{1}{c|}{35.24} & \multicolumn{1}{c|}{0.76} & \multicolumn{1}{c|}{0.80} & 0.76 \\ \hline
\end{tabular}
\end{table*}

Table~\ref{tab:NLP Models} presents the averaged results across folds, obtained after excluding 'None' labels from each annotator.

As shown in Table~\ref{tab:NLP Models}, both DistilBERT and MentalRoBERTa demonstrated varying performance across different annotators. Llama3 and MentaLLaMA achieved the highest training accuracy ($95.85$\% and $94.15$\% for DistilBERT, $91.02$\% and $92.35$\% for MentalRoBERTa), but exhibited significant over-fitting or sensitivity to some specific disorders it was trained on, as evidenced by the substantial drop in validation accuracy ($70.31$\% and $59.21$\% for DistilBERT, $69.42$\% and $60.64$\% for MentalRoBERTa). Dictionary matching offered consistent and stable performance, particularly in terms of validation accuracy ($86.25$\% with DistilBERT and $82.73$\% with MentalRoBERTa) but was limited by its reliance on explicit keyword matching. Fig.~\ref{fig:condition_counts} highlights the class imbalance in the dictionary matching dataset, with a high concentration of `\textit{None}' labels ($50.5$\%), followed by \textit{Depression} ($30.26$\%) and \textit{Anxiety} ($10.93$\%), which might have contributed to its relatively strong performance due to potential biases towards majority classes. However, this approach is inherently limited in capturing subtle emotional cues and contextual details, which constrains its ability to fully comprehend the nuances of the posts. The F1 Scores for Llama3 ($0.88$ for both) and dictionary Matching ($0.87$ for MentalRoBERTa and $0.85$ for DistilBERT) were the highest among the annotators, suggesting that these models performed better in maintaining a balance between precision and recall. While dictionary matching offers consistency and accuracy, Llama3 and MentaLLaMA demonstrated potential for capturing nuanced language patterns, although with a higher risk of over-fitting. Samantha-Mistral, despite its overall low performance, provided valuable insights into areas requiring further model refinement. Overall, the DistilBERT model, when fine-tuned with Llama3, provided the best generalization across the different annotators.

\subsubsection*{Analysis of labels}
\label{subsecr:labels}
\begin{figure}[t]
    \centering
     \vspace{-0.2cm}
    \includegraphics[width = \linewidth]{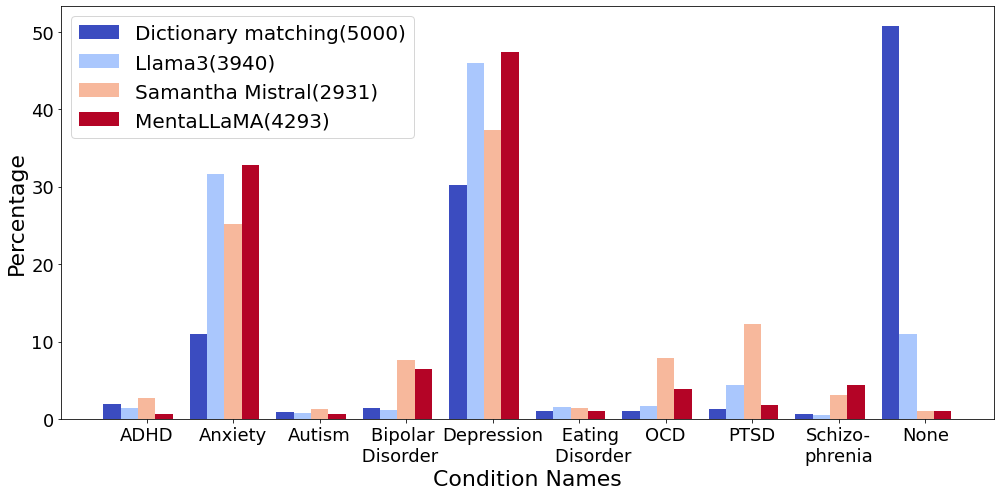}
    \vspace{-0.6cm}
    \caption{Normalized condition counts for various annotators}
    \label{fig:condition_counts}
\end{figure}

\begin{figure}[t]
    \centering
     \vspace{-0.2cm}
    \includegraphics[width = \linewidth]{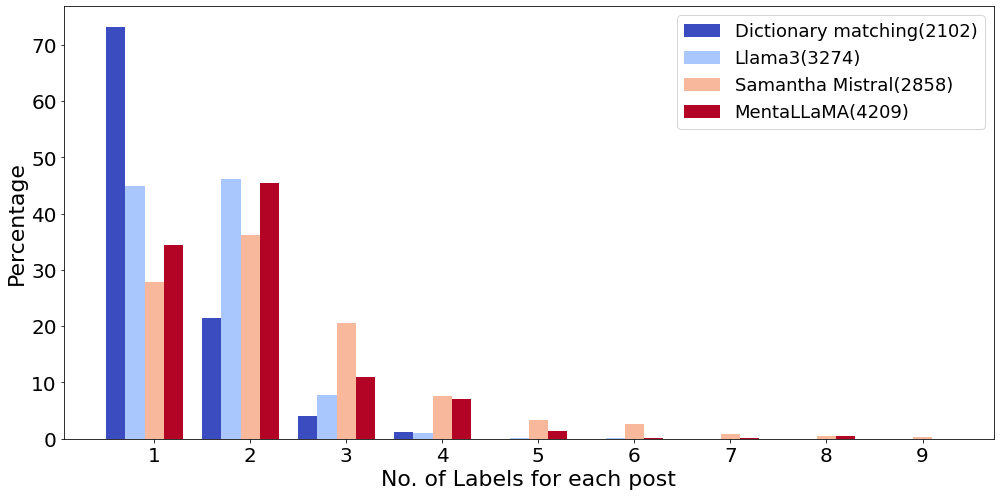}
    \vspace{-0.6cm}
    \caption{Number of labels for each post for various annotators}
    \label{fig:label_counts}
\end{figure}

The distribution of different labels across annotators is shown in Fig.~\ref{fig:condition_counts} while the distribution of label counts per post for each annotator is presented in Fig.~\ref{fig:label_counts}.

From Fig.~\ref{fig:condition_counts}, we observe that \textit{Depression} and \textit{Anxiety} were the most frequently assigned labels across all annotators, with a pronounced dominance in the dictionary-based method. This suggests these disorders are either highly prevalent or more easily detectable using keyword-based approaches. Llama3 and MentaLLaMA demonstrated a more balanced distribution of labels, indicating better detection of less common disorders. A substantial portion of posts were labeled as `\textit{None}' by the dictionary-based method ($50.74\%$), implying either a conservative labeling approach or the absence of relevant keywords in the text whereas for Llama3 it was $10.93$\%, which consisted mostly of the suicidal/self-harm cases. For MentaLLaMA and Samantha Mistral, it turned out to be less than $1$\%.

Fig.~\ref{fig:label_counts} shows that most posts receive a single label for dictionary-based annotation which turned out to be more conservative. In contrast, Large Language Models (LLMs) tended to assign multiple labels, suggesting a greater capacity to identify overlapping disorders. Across all annotators, the frequency of posts decreased with increasing label count. While the dictionary-based method identified up to six labels per post, LLMs detected up to nine, indicating their enhanced ability to recognize multiple disorders.

\subsection{Severity Classification}
\label{subsecr:severity}
Following the pre-processing step, we used various LLMs to categorize the severity level of each post into three classes -- \textit{mild}, \textit{moderate}, and \textit{severe}, as described in Sec.~\ref{subsecm:severity}.
All three LLMs classified the majority of the posts into the \textit{Severe} category, with Samantha-Mistral classifying no posts as \textit{Mild} (see Table.~\ref{table:severity}).
The LLMs identified and interpreted indicators of distress, disorder severity, and urgency in mental health discussions, enabling the categorization of posts by disorder intensity. This allows for targeted analysis and appropriate therapeutic recommendations, making it a key component in understanding and addressing varying mental health challenges.
\begin{table}[h]
\caption{Analysis of Severity Level Classification}
	\label{table:severity}
	\begin{center}
	    \begin{tabular}{|p{3.3cm}|r|r|r|}
        \hline
        \textbf{Model Name} & \textbf{Mild} & \textbf{Moderate} & \textbf{Severe} \\ 
        \hline
        Llam3 & 1.5\% & 17.4\% & 81.1\% \\ 
        \hline
        MentaLLaMA & 7.4\% & 10.3\% & 82.3\% \\ 
        \hline
        Samantha-Mistral & 0\% & 3\% & 97\% \\ 
        \hline
        \end{tabular}
	\end{center}
\end{table}

\subsection{Recommendation Module}
\label{subsecr:recomm}
Our queries on LLMs produced a list of recommended therapies. Highly recommended therapies include Cognitive Behavioral Therapy (CBT), Dialectical Behavior Therapy (DBT), Acceptance and Commitment Therapy (ACT), Psychodynamic Therapy (PT), Mindfulness-Based Stress Reduction (MBSR), Mindfulness-Based Cognitive Therapy (MBCT), Interpersonal Therapy (IPT), Exposure Therapy (ET), Motivational Interviewing (MI), and Family Therapy (FT). Other Therapy (OT) groups all the less frequent therapies, while No Therapy (NT) is suggested when LLMs recommend users to seek professional help, particularly in `extreme' cases where it detects a tendency towards self-harm or suicidal thoughts.

CBT is the most recommended therapy for all three LLMs as a result of its adaptability and broad applicability. However, their second most recommended differ: Llama3 prefers ACT, MentaLLaMA prefers DBT, whereas Samantha-Mistral focuses on less frequent therapies.

We mapped the disorders detected in posts to the therapies recommended in them. 
We observed that ACT is more recommended than CBT by Llama3 for disorders like anxiety and depression. This finding is significant, as CBT has traditionally been considered the standard approach for these problems. However, Llama3's preference for ACT suggests a greater emphasis on mindfulness, acceptance, and commitment to personal values, highlighting the model's inclination towards therapies that focus on holistic mental well-being.

We have also asked the LLMs to recommend behavior changes for these posts (see Sec.~\ref{subsecm:recom}). We observe that LLMs are suggesting actionable tasks to the user to improve their mental health. The LLMs provide behavior change suggestions aimed at practical daily actions for users to incorporate into their lives. These include mindfulness techniques, healthy routines, and coping mechanisms for mental health issues. The suggestions focus on small, manageable tasks to foster a positive outlook and empower individuals to take control of their mental well-being. This approach addresses immediate emotional needs and supports long-term mental health, laying a foundation for sustained personal growth and self-care.

\section{Time complexity of different LLMs}
\label{sec:complexity}
The evaluation of three LLMs -- Llama3, MentalLaMA, and Samantha-Mistral -- was conducted on a TP100 GPU with $15$ GB memory. Llama3, the largest and most parameter-heavy model, took $40$ minutes to process $100$ samples. MentaLLaMA completed the same task in $19$ minutes, while Samantha-Mistral took $100$ seconds. However, Samantha-Mistral’s speed comes with reduced effectiveness, making it less suitable for complex tasks. In summary, Llama3 excels in capability but demands resources, MentalLaMA balances speed and accuracy, and Samantha-Mistral, though fast, struggles with deeper analysis.

\section{Discussions}
In this paper, we present the comprehensive framework for the evaluation of multiple pre-trained NLP models and LLMs involving binary, multi-class, multi-label classification, and recommendation tasks from social media discussions related to mental health. 
We observed that while the dictionary-matching approach is more stable, it lacked the depth required to identify disorders based on the context and sentiment of the posts, which could limit its effectiveness in complex scenarios. This can be further fine-tuned by including stylistic features, emotion, and Empath libraries to make it more reliable. Our proposed framework is not only robust in identifying a wide range of mental health disorders but also versatile enough to provide tailored recommendations for therapy and behavior changes. Although we present results for a sample dataset, the framework can be scaled up for much larger datasets, making it a valuable tool for large-scale mental health assessment and recommendations.

Moreover, careful prompt engineering of LLMs has proven effective in generating comprehensive insights, including disorder identification, severity assessment, recommendation of therapy and behavior changes. 
We observed that Llama3 outperformed the other two LLMs in disorder identification. For other tasks, MentaLLaMA delivered better results. However, Samantha-Mistral, although being the fastest, has not yet met the performance standards for these tasks. 

Ethical considerations are essential throughout this process. While the framework prioritizes accurate analysis, language models are designed to handle sensitive content with care, often refusing to generate output related to suicide or self-harm to uphold ethical standards and ensure user safety. 

Our framework has the potential to contribute significantly to the field of mental health informatics. By developing more accurate methods for automatically identifying mental health disorders from social media data, we can potentially enable earlier interventions and improve the lives of individuals struggling with these problems. These insights could be instrumental for mental health practitioners, offering them a new resource for monitoring and intervening in patient care. Additionally, the methodology has the capability to supplement a digital well-being platform, providing personalized mental health support directly to users.

\bibliography{WI296}
\bibliographystyle{IEEEtran}

\end{document}